\documentclass[a4paper,11pt]{article}
\pdfoutput=1 

\usepackage{jheppub} 

\usepackage[T1]{fontenc} 
\usepackage{graphicx}
\usepackage{dcolumn}
\usepackage{bm}
\usepackage{slashed}
\usepackage{natbib}

\title{Sum rules and the DGLAP evolution in presence of $U(1)_B$ anomaly}

\author[a]{Thitipat Sainapha}

\affiliation[a]{Chulalongkorn University, Bangkok, Thailand}

\emailAdd{thitipat22552@gmail.com}
\emailAdd{Thitipat.S@student.chula.ac.th}

\abstract{The possibility to modify both quarks sum rules and the DGLAP evolution equation are investigated. In presence of anomalous baryon global symmetry, the quark sum rules are no longer the same as usual. In addition, the DGLAP equation is also changed to be self-consistent with modified sum rules. This way of modification turns out to satisfy the Sakharov's baryogenesis requirements. We also assume the existence of the factorization theorem in relevant scale to maintain the naive parton model. Consequently, if this is presumably going to be true, there might be a new powerful constraint for constructing the beyond standard model physics.}

\begin{document} 
\maketitle
\flushbottom

\section{Introduction}
\indent\indent Baryogenesis, the particles-anti-particles asymmetry, is one of the most interesting problems in the standard model of particle physics for a half of a century. It is the motivation for finding beyond the standard model physics. In this paper, we will explore the possibility of contruction of the baryogenesis through the theory that baryon numbers conservation can be violated by quantum correction, particularly called an anomalous global baryon symmetry. The study of quantum
anomalies can be found in many literatures \cite{Fujikawa1980,Adler1969,Bell1969,Atiyah1968}. Global anomalies can have physical consequences without destroying the theory unlike anomalous gauge symmetries. Despite the assumption that baryon numbers anomaly exists in some high energy scale such as GUT scale, the quark sum rules will be changed accordingly. Consider in high temperature and high density regime, the process involving proton can be studied by the parton model by assuming the existence of the factorization theorem. The quark sum rules are the constraints to force the parton distribution function to have probabilistic interpretation. Additionally, the DGLAP evolution equation, after Dokshitzer, Gribov, Lipatov, Altarelli and Parisi, determining the behavior of parton distribution function at any scale \cite{Altarelli1977,Dokshitzer:1977sg,Gribov:1972ri} have to be modified. We expect that all of these facts must respect the condition for existence of baryogenesis known as the Sakharov's conditions \cite{Sakharov1991} states that the theory which can be used to explain the baryogenesis must have 
\begin{enumerate}
    \item violation of baryon number conservation.
    \item charge-parity violation (CP violation).
    \item thermal non-equilibrium phase transition.
\end{enumerate} 
\indent The outline of this paper will be presented as follows\\
\indent In Sec. II we will review the derivation of anomalies by determining the measure changing under chiral symmetry transformation within path integral formalism. Moreover, we will explore the properties of baryon numbers anomaly such as general perspective, Atiyah-Singer index theorem, and so on. Moreover, we suggest the way to modify the anomalous term to be consistent to the observation in the real world. All of these are the preparation for the next section.\\
\indent In Sec. III we shall use the results obtained from Sec. II to relate with naive parton model. To study the possible way to modify the quarks sum rules and the DGLAP evolution equation which is the main point of this work.

\section{Derivation of anomalies}
\indent\indent The most elegant and attractive way to derive anomalies is the path integral approach firstly proposed by Fujikawa \cite{Fujikawa1980}. In this section, we will give a quick review on this procedure. First of all, let's begin with massless QCD Lagrangian expressed as following:
\begin{equation}
    \mathcal{L}=-\frac{1}{2}Tr\{F_{\mu\nu}F^{\mu\nu}\}+i\bar\psi\slashed{D} \psi.
\end{equation}
where Tr traces over color indices. A non-Abelian field strength tensor is defined as $F_{\mu\nu}\equiv \partial_\mu A_\nu -\partial_\nu A_\mu -i[A_\mu,A_\nu]$. This Lagrangian is invariant under chiral rotation $\psi\rightarrow e^{i\gamma^5\beta}\psi$ with fifth gamma $\gamma^5\equiv i\gamma^0\gamma^1\gamma^2\gamma^3$, then we can define the axial-vector conserved current associated to this symmetry. Unfortunately, in quantum level, this Noether's current is no longer conserved. To see that more clearly, it is more convenient to work in Euclidean space so we do analytic continuation by Wick rotation $z^0\rightarrow -iz^4$, $A_0\rightarrow iA_4$. The advantage of doing this is the Dirac differential operator $\slashed{D}$ becomes Hermitian. Following these logic, we can define a Dirac eigenfunction such that
\begin{equation}
    \slashed{D}\varphi_n(z)=\lambda_n\varphi_n(z),
\end{equation}
satisfying normalization condition and completeness relation
\begin{equation}
    \int d^4z \varphi_m^\dagger(z) \varphi_n(z) = \delta_{mn},\; \sum_n \varphi_n(z)\varphi_n(y)^\dagger =\delta^4(z-y)
\end{equation}
Defining this eigenfunction will be helpful later. For local chiral transformation, the Lagrangian can be changed. In addition, the measure of path integration is also changed
\begin{equation}
    D\bar\psi D\psi \rightarrow D\bar\psi D\psi \;det[e^{2i\gamma^5\beta(z)}]^{-1}.
\end{equation}
Using identity $det(e^A)=e^{trA}$ where trace operator is tracing over all possible Hilbert space basis. Thus, the total change of integral measure is then
\begin{equation}
    \begin{array}{cc}
         =&exp\left(-2iTr\int d^4z <z|\gamma^5\beta(x)|z>\right)  \\
         =&exp\left(-2iTr\int d^4z \sum_n \varphi^\dagger_n(z)\gamma^5\beta(z)\varphi_n(z)\right),
    \end{array}
\end{equation}
where in the last step, we have used the completeness relation of Dirac eigenfunction (2.3) and Tr here is not Hilbert space basis trace but Dirac trace over gamma matrices. Note that the result (2.5) goes as $0\times \infty$ since $Tr\gamma^5=0$ and $\sum_n\varphi_n\varphi^\dagger_n=\delta^4(0)\sim \infty$. We, therefore, need to regulate the result by cutting off by some regulator with specific asymptotic behaviors. The derivation of equation (2.6) will be discussed in appendix A of which the final result takes form
\begin{equation}
    det(e^{-2i\gamma^5\beta(z)})=exp\left[i\int d^4z \left(\beta(z)\frac{g^2}{8\pi^2}Tr\{F_{\mu\nu}\tilde{F}^{\mu\nu}\}\right)\right],
\end{equation}
with color trace Tr and dual field strength tensor $\tilde{F}^{\mu\nu}\equiv \frac{1}{2}\epsilon^{\mu\nu\alpha\beta}F_{\alpha\beta}$. On the other hand, the Lagrangian (2.1) is also transformed under local chiral rotation as 
\begin{equation}
    \mathcal{L}\rightarrow\mathcal{L}-J^{5\mu}\partial_\mu\beta(z).
\end{equation}
However, the all possible path integration must be invariant according to the Schwinger's statement of quantum action principle. We obtain
\begin{equation}
    <\partial_\mu J^{5\mu}>=-\frac{g^2}{8\pi^2}Tr\{F_{\mu\nu}\tilde{F}^{\mu\nu}\}.
\end{equation}
As we have mentioned before that classically conserved current can be broken by quantum correction. Roughly, the symmetry is said to be anomalous. In general, all gauge anomalies must be cancelled perfectly in gauge theories because the existence of anomalous terms will violate the Ward-Takahashi identity which is the requirement of unitarity and renormalizability of the theory. Global symmetries have nothing to do with the Ward identity, then global symmetry can be safely anomalous. Thus, only anomalous global symmetry can have physical implications.
\subsection{Anomalous baryon global symmetry}
\indent\indent Alternatively, the quantum anomalies can be derived through the direct calculation of triangle diagram with currents insertion to each vertex and fermions running around the triangle loop. (Such a derivation can be found in several literature \cite{Adler1969},\cite{Bell1969}). This alternative way to derive anomalies is somehow meaningful. One can calculate the insertion of $U(1)_BSU(2)SU(2)$ currents where $U(1)_B$ is baryon global symmetry. One can derive using the same Fujikawa method to obtain precisely
\begin{equation}
    \partial_\mu J^\mu_B=\frac{3g_w^2}{8\pi^2}Tr\{F_{\mu\nu}\tilde{F}^{\mu\nu}\}.
\end{equation}
The particular factor 3 comes from the number of the standard model's generations of quarks. Note that the right hand side of (2.9) can be rewritten as the total derivative of something else. Explicitly,
\begin{equation}
    Tr\{F_{\mu\nu}\tilde{F}^{\mu\nu}\}=2\partial_\mu(Tr\{\epsilon^{\mu\nu\alpha\beta}(A_\nu\partial_\alpha A_\beta-\frac{2i}{3}A_\nu A_\alpha A_\beta)\}).
\end{equation}
We conventionally define the expression inside the trace operation in the right hand side of (2.10) as
\begin{equation}
    K^\mu\equiv \epsilon^{\mu\nu\alpha\beta}(A_\nu\partial_\alpha A_\beta-\frac{2i}{3}A_\nu A_\alpha A_\beta)
\end{equation}
which is called the Chern-Simon current (CS current). Finally, we can easily rewrite the anomalous baryon number conservation as follows
\begin{equation}
    \begin{array}{ccc}
       \partial_\mu J^\mu_B  &=&  \frac{3g_w^2}{4\pi^2}\partial_\mu K^\mu, \\
       J^\mu_B  &=& \frac{3g_w^2}{4\pi^2} K^\mu+J^\mu_0.
    \end{array}
\end{equation}
$J^\mu_0$ appears to be an integration constant which is not necessary to be zero. Consider only the time component of the current, $\mu=0$, integrating over total spacial coordinate. We have
\begin{equation}
    B=\frac{3g_w^2}{4\pi^2}Q_{CS}+B_0.
\end{equation}
B is total numbers of baryon in the theory, $B_0$ is non-anomalous baryon numbers and $Q_{CS}\equiv\int d^3x K^0$ the Chern-Simon charge. Express the total baryon numbers in terms of numbers of quarks and anti-quarks. In particular, $B=\frac{1}{3}(n_q-n_{\bar{q}})$, we get
\begin{equation}
    \tilde{n}_q-\tilde{n}_{\bar{q}}=\frac{9g_w^2}{4\pi^2}Q_{CS}.
\end{equation}
Here we have already absorbed the initial baryon numbers into $\tilde{n}$. The reason why we write the relation this way is to make it be precisely reminiscent of the famous Atiyah-Singer index theorem \cite{Atiyah1968}. Thus, we can think of $\tilde{n}_q$ and $\tilde{n}_{\bar{q}}$ as the total number of topologically invariant zero modes of some elliptic operator. Presumably, if it is truly Atiyah-Singer theorem, this implies that one might be able to compute quark-anti-quark imbalance by the computation of the index of these operator. \\
\indent Honestly, we will end this section by phenomenological modifying this model slightly. In the range of reachable energy scale, we have not observed baryon numbers anomaly. Thus, we modify the right hand side term of (2.14) by adding Heaviside step function as follows
\begin{equation}
    \tilde{n}_q-\tilde{n}_{\bar{q}}=\frac{9g_w^2}{4\pi^2}Q_{CS}\theta(ln\mu-ln\Lambda).
\end{equation}
$\Lambda$ denoted the energy scale existing the anomalous of $U(1)_B$ symmetry, e.g. the grand unification scale $\sim 10^{16}$ GeV or some supersymmetric model's scale. The reason we write expression inside Heaviside step function in term of logarithm will become clear later.\\
\section{Relates to parton model}
\indent\indent From now on, we will, in relevant energy scale, assume the existence of the factorization theorem such that any high energy process involving proton can be parametrized by universal structure called parton \cite{Bjorken1969}. There is an important quantity to study the parton model, which is the classical probability for finding parton of species i in precise energy scale $\mu$ and Bjorken factor $x$, also known as the parton distribution function (PDF) denoted by $f_i(x,\mu)$.
\\
\indent The point is to interpret PDFs as classical probability, they need to satisfy the constraint called sum rules. One of the important sum rules is 
\begin{equation}
    \int_0^1 dx f_i(x,\mu)=n_i(\mu),
\end{equation}
this means physically that the summation over all possible Bjorken of PDF of species i yields the total numbers of parton of that species. Hence, the total baryon number sum rule takes form
\begin{equation}
    B=\sum_q\frac{1}{3}\int dx (f_q(x,\mu)-f_{\bar{q}}(x,\mu)).
\end{equation}
where the summation is summing over all flavors of quarks that are hadronized into baryons. Note that we only consider the case that quarks are hadronizable or high enough density. The another way to think about this sum rule is this rule is directly proportional to Gross-Llewellyn-smith sum rule \cite{hep-ph/9604210,GrossLlewellyn} up to additive constant, specifically unmodified strangeness quantum number. This sum rule can be determined as the form factor $F_3$ measured from neutrino deep inelastic scattering with equal neutron-proton nuclear target. In particular, the experiment must perform at center of mass energy around baryon number violation scale.
\\
\indent Back to our consideration, substituting the result from last section (2.13) with the modification (2.15) and re-write PDFs as the summation of non-anomalous PDFs and anomalous contribution of PDFs, $\tilde{f}$ says. We finally have
\begin{equation}
    \int dx \; \sum_{q,\bar{q}}(\tilde{f}_q(x,\mu)-\tilde{f}_{\bar{q}}(x,\mu))=\frac{9g_w^2}{4\pi^2}Q_{CS}\theta\left(ln\frac{\mu}{\Lambda}\right).
\end{equation}
This modifies the quark sum rules from original sum rules to the system containing baryon numbers anomaly. Normally, we assume that partons are weakly interacting at high energy level which is the indirect consequence of asymptotic freedom \cite{Gross:1973ju}. Thus, PDFs are UV stable or independent of energy scale $\mu$, this phenomenon is also known as Bjorken scaling \cite{Bjorken1969scaling}. However, in reality, PDFs are $\mu$-dependent suppressed by large logarithm. To resummation the large logarithm contribution, we allow them to run as change of energy level. The renormalization group equation demonstrating the running of PDFs are widely known as the DGLAP equation (See also \cite{Altarelli1977,Dokshitzer:1977sg,Gribov:1972ri})
\begin{equation}
    \mu\frac{d}{d\mu}f_i(x,\mu)=\frac{\alpha_s}{\pi}\int^1_x\frac{d\xi}{\xi}f_i(\xi,\mu)P_{qq}\left(\frac{x}{\xi}\right),
\end{equation}
where $\alpha_s$ be fine-structure constant of strong interaction $\equiv \frac{g^2}{4\pi}$ and $P_{qq}\left(\frac{x}{\xi}\right)$ so-called the DGLAP splitting function. These splitting function satisfies
\begin{equation}
    \int^1_0dy\; P_{qq}(y)=0,
\end{equation}
which we will use this identity soon. Now let's see this behavior in anomalous baryon system by taking derivative respects to $ln\mu$ into (18). To get
\begin{equation}
    \sum_{q,\bar{q}}\int^1_0 dx \int^1_x \frac{d\xi}{\xi}P_{qq}\left(\frac{x}{\xi}\right)-(q\leftrightarrow \bar{q})=9\frac{Z^{-2}_{gw}g^2_w}{g^2_s}Q_{CS}\delta\left(ln\frac{\mu}{\Lambda}\right)
\end{equation}
where $\frac{d\theta(x)}{dx}=\delta(x)$ has been used in the last step and $Z_{gw}$ is renormalization factor for renormalized weak coupling. Note that $Q_{CS}$ is expected to be scale-independent since it is the operator appearing in Lagrangian of specific energy scale generally containing infinities in it. We can also write it as the renormalized CS charge which does not have to do specifically in this paper anyway. Luckily, the useful fact about renormalized quantity in anomalous field theory is only 1-loop renormalization factors (Can be found in many standard books, e.g. \cite{QFTSchwartz}) are needed since anomalies are proven to be 1-loop exact \cite{Adler1969}. Let's try to calculate the renormalized first term in CS charge
\begin{equation}
    g^2_0A_0\partial A_0\rightarrow (Z_1^2Z^{-2}_2Z_3)g^2(Z_3)A\partial A=\frac{Z^2_1}{Z_2^2}g^2A\partial A.
\end{equation}
1-loop renormalization factors $Z_1\not= Z_2$ in general non-Abelian gauge theories (One way to think why $Z_1\not= Z_2$ is because $Z_1$ generally is the renormalization factor for 3-point vertex, while $Z_2$ is the renormalization factor for gauge field renormalization, the non-equality of these two factors means that 3-point vertex received radiative correction always happened in non-Abelian theories). Thus, we can conclude that the current is running in presence of anomalies. The particular reason is the unrenormalized property of the current in usual quantum field theories is the attractive consequence of the Ward-Takahashi identity violated by anomalies (Remember that Ward-Takahashi identity can be thought as momentum space version of conservation of current). This also implies physically that the total number of particles minus anti-particles is scale-dependent as we expected.\\
\indent Further note about equation (3.7), at leading order of the renormalized couplings, $\frac{\alpha_w}{\alpha_s}$ ratio is strongly suppressed in low energy level whereas becomes those of order $\sim \mathcal{O}(1)$ at around the GUT scale. Thus, even though some SUSY model can have anomalous baryon symmetry, this correction will become significant around the GUT scale only.\\
\indent Unfortunately, this has not completed yet. To see the problem, let's apply the condition (3.5) in left hand side of (3.6). Each term vanishes independently so the final result will be not consistent with itself. This implies interestingly that the DGLAP evolution equation (3.4) must be also modified consequently. These presumably becomes
\begin{equation}
         \mu\frac{d}{d\mu}f_i(x,\mu)=\frac{\alpha_s}{\pi}\int^1_x\frac{d\xi}{\xi}f_i(\xi,\mu)P_{qq}\left(\frac{x}{\xi}\right)+\frac{9\alpha_w}{\alpha_s}a_qQ_{CS}\delta\left(ln\frac{\mu}{\Lambda}\right).
\end{equation}
where renormalization factor for current have already been absorbed into renormalized CS charge conventionally. On the other hand, the modified DGLAP evolution equation for anti-quark has the same form with interchanging $q\leftrightarrow\bar{q}$. $a_q$ denoted the weight factor of anomalous term for quark PDF evolution, generally assuming that those can be different from those of anti-quark. However, they have to satisfy the constraint $\sum_{q,\bar{q}}a_q-a_{\bar{q}}=1$ for self-consistency.\\
\indent The final remark can be seen explicitly by not applying the DGLAP equation, alternatively writing (3.6) to be
\begin{equation}
    \mu\frac{d}{d\mu}(n_q-n_{\bar{q}})=\frac{9g_w^2}{4\pi^2}Q_{CS}\delta\left(ln\frac{\mu}{\Lambda}\right).
\end{equation}
The relation (3.9) is very meaningful. At the energy scale $\mu\sim T=\Lambda$ up to the Boltzmann constant, the running of the difference between quark numbers and anti-quark numbers blow up. The physical meaning is there exists the first-order thermal phase transition at these specific scale. Additionally, the existence of CS conserved charge violates both Baryon number conservation and discrete charge-parity (CP) symmetry. All of this properties are requirements of Baryogenesis proposed by Sakharov \cite{Sakharov1991}. The explicit phase transition might probably be able to be derived in the context of finite temperature and finite density quantum field theory.
\section{Conclusions}
\indent\indent The fact that $U(1)_B$ symmetry is anomalous in some high energy scale leads us to possibly modify the quark sum rules. This also forces us to modify the DGLAP evolution equation. This possible way to generalize is surprisingly to be consistent to all of the Sakharov's requirements for particles-anti-particles imbalance or Baryogenesis. This model may be able to be verified by determining the anomalous contribution in $F_3$ form factor, performing the neutrino deep inelastic scattering around GUT center of mass energy scale. However, this model is needed for further studying especially in context of finite temperature and finite density quantum field theory. To end this, note that the results obtained in this paper will be consistent if and only if the factorization theorem holds at the relevant energy level. The reason to propose this way of modification of the quarks sum rules and the DGLAP evolution equation, as already mentioned in the abstract, is it might be able to become a constraint for constructing beyond the standard model physics.

\appendix
\section{Regularization of integral measure}
\indent\indent Recall that the integral measure changing under chiral rotation is ill-behaved. We need to regulate by some smooth function $f$ with specific asymptotic behaviors $f(0)=1$, $f(\infty)=0$. Originally, Fujikawa chose the regulator to be the Gaussian regulator $e^{-(\lambda_n/M)^2}$ where M is cut off scale much larger than $\lambda_n$. Consider
\begin{equation}
    \begin{array}{ccc}
       \displaystyle{Tr\left(\int d^4z \sum_n \varphi_n^\dagger(z)\gamma^5\varphi_n(z)\right)}
       &=& \displaystyle{\lim_{M\rightarrow\infty}}Tr\left(\int d^4z \sum_n \varphi_n^\dagger(z)\gamma^5e^{-\left(\frac{\lambda_n}{M}\right)^2}\varphi_n(z)\right) \\
        &=&\displaystyle{\lim_{M\rightarrow\infty}}Tr\left(\int d^4z \sum_n \varphi_n^\dagger(z)\gamma^5e^{-\left(\frac{\slashed{D}}{M}\right)^2}\varphi_n(z)\right).
    \end{array}
\end{equation}
To evaluate this, we write the Dirac operator as
\begin{equation}
    \slashed{D}^2=D^2-\frac{g}{2}\sigma_{\mu\nu}F^{\mu\nu}.
\end{equation}
We can evaluate the Dirac trace of second term of the right hand side of (A.2) by using the fact that the trace over fifth-gamma with other gamma matrices unless with 4 gammas. Thus, the leading order non-vanishing contribution comes from $\mathcal{O}(M^{-4})$. Using the identity $\frac{1}{2}\{\sigma^{\mu\nu},\sigma^{\alpha\beta}\}=g^{\mu\alpha}g^{\nu\beta}-g^{\mu\beta}g^{\nu\alpha}+i\gamma^5\epsilon^{\mu\nu\alpha\beta}$ so that $(\sigma_{\mu\nu}F^{\mu\nu})^2=2F^{\mu\nu}F^{\mu\nu}+i\gamma^5\epsilon^{\mu\nu\alpha\beta}F_{\mu\nu}F_{\alpha\beta}$. Combining everything we have known to obtain
\begin{equation}
    Tr\left(\int d^4z \sum_n \varphi_n^\dagger(z)\gamma^5\varphi_n(z)\right)=-g^2F_{\mu\nu}\tilde{F}^{\mu\nu}\lim_{M\rightarrow\infty}\frac{1}{M^4}\sum_n\varphi_n^\dagger(z)e^{-\left(\frac{p}{M}\right)^2}\varphi_n(z).
\end{equation}
Following these steps, Fourier transforming the result, we have
\begin{equation}
   Tr\left(\int d^4z \sum_n \varphi_n^\dagger(z)\gamma^5\varphi_n(z)\right) =-g^2F_{\mu\nu}\tilde{F}^{\mu\nu}\lim_{M\rightarrow\infty}\frac{1}{M^4}\int \frac{d^4k}{(2\pi)^4}e^{-\left(\frac{k}{M}\right)^2}.
\end{equation}
Finally, we just need to perform the integral; fortunately, the integral is simple to evaluate because we have already analytically continuation to Euclidean space
\begin{equation}
    \frac{1}{M^4}\int \frac{d^4k}{(2\pi)^4}e^{-\left(\frac{k}{M}\right)^2}=\frac{1}{16\pi^2}.
\end{equation}
Plugging the result (A.5) into (A.4), we will end up with desired result (2.6) have been claimed in section II.

\acknowledgments

\indent\indent I would like to thank Karankorn Kritsarunont and Kittipat Hiraprayoonpong for very helpful discussions. Also thank Inori Minase, Sayaka Yamamoto and idols in 26ji no masquerade band for sweet songs and smiles when I'm feeling down about my life.




\end{document}